\documentclass[aps,groupedaddress,amsmath,amssymb,showpacs,prl,twocolumn]{revtex4}
\usepackage{amssymb,amsmath}
\usepackage{graphicx}

\begin{document}

\title{Band gap corrections for molecules and solids using Koopmans' theorem and Wannier functions}

\author{Jie Ma}
\affiliation{Joint Center for Artificial Photosynthesis and Materials Sciences Division, Lawrence Berkeley National Laboratory, Berkeley, California 94720, USA}

\author{Lin-Wang Wang}
\email{lwwang@lbl.gov}
\affiliation{Joint Center for Artificial Photosynthesis and Materials Sciences Division, Lawrence Berkeley National Laboratory, Berkeley, California 94720, USA}

\begin{abstract}
We have proposed a method for correcting the Kohn-Sham eigen energies in the density functional theory (DFT) based on the Koopmans' theorem using Wannier functions. The method provides a general approach applicable for molecules and solids for electronic structure calculations. It does not have any adjustable parameters and the computational cost is at the DFT level. For solids, the calculated eigen energies agree well with the experiments for not only the band gaps, but also the energies of other valence and conduction bands. For molecules, the calculated eigen energies agree well with the experimental ionization potentials and electron affinities, and show better trends than the traditional Delta-self-consistent-field ($\Delta$SCF) results.
\end{abstract}


\maketitle

Density functional theory (DFT)~\cite{HK} is the main working horse for material simulations, especially for ground-state properties such as atomic structures and binding energies. However, it is well known that the DFT, in particular the Kohn-Sham eigen energy~\cite{KS}, significantly underestimates band gaps. This is related to the lack of derivative discontinuity in the exchange-correlation (XC) potential when the total number of electrons crosses an integer point~\cite{discontinuity1,discontinuity2}. Over the years, various methods have been developed to overcome this deficiency. One popular approach is the hybrid functional~\cite{hyb1,hyb2,hyb3}, which mixes the exact exchange with local/semilocal XC functionals. Although widely successful, these methods depend on the mixing parameters and are computationally more expensive than local/semilocal functionals such as the local density approximation (LDA) or generalized gradient approximation. An even higher-level method is the GW and the related random-phase approximation (RPA)~\cite{GW,RPA}, but the high computational cost of RPA makes it only applicable to small molecules~\cite{RPA2}. There exists another approach to correct the DFT and its related Kohn-Sham Hamiltonian: the Koopmans' theorem~\cite{koopman-note,koop1,koop2,koop3,koop4,koop5,yang,koop6}. Under the Koopmans' theorem, the total energy $E(n)$ as a function of the continuous number of electrons $n$ should be a straight line between two integer points~\cite{koopman-note,Perdew}.
However, the LDA $E(n)$ curve is convex~\cite{yang2,yang1}. According to the Janak's theorem, for a $N$-electron system, $\epsilon_{\rm VBM}=\frac{\partial E(n)}{\partial n}|_{n\rightarrow N^-}$, $\epsilon_{\rm CBM}=\frac{\partial E(n)}{\partial n}|_{n\rightarrow N^+}$ with $\epsilon_i$ being the eigen energies of the valence band maximum (VBM) and conduction band minimum (CBM)~\cite{janak,yang1,yang2}, so the convexity of the LDA energy leads to an underestimation of the band gap $\epsilon_{\rm CBM}-\epsilon_{\rm VBM}$.
One explicit way to enforce the straight-line condition is to modify the LDA total energy to
\begin{equation}
E=E_{\rm LDA}+\sum_l \tilde{E}_l(s_l) \label{newE}\,,~s_l\in[0,1]
\end{equation}
with $\tilde{E}_l(s_l)=[E_l(N\pm 1)-E(N)]\times s_l-[E_l(N\pm s_l)-E(N)]$ ($+$ for adding electrons to unoccupied orbitals, and $-$ for removing electrons from occupied orbitals). $E_l(N\pm s_l)$ is the self-consistent-field LDA energy after adding/removing $s_l$ electrons on the $\phi_l$ orbital. Note $\tilde{E}_l(s_l)=0$ for $s_l=0$ or 1. Taking the variational minimum of $E$ with respect to $\psi_i$ (here $s_l=|\langle\phi_l|\psi_i\rangle|^2$ is the projection of $\psi_i$ on $\phi_l$), we obtain a modified Kohn-Sham equation
\begin{equation}
[H_{\rm LDA}+ \sum_l \lambda_l |\phi_l\rangle\langle\phi_l|] |\psi_i\rangle = \epsilon_i |\psi_i\rangle \label{newH}
\end{equation}
with $\lambda_l = \frac{\partial \tilde{E}_l(s_l)}{\partial s_l}|_{s_l=0}$, providing corrections to eigen energies.

Using eigen orbitals as $\phi_l$, the eigen energies in Eq.(\ref{newH}) work successfully for atoms and small molecules~\cite{koop3,yang,koop4,koop5}, because the straight-line condition makes the VBM and CBM eigen energies equal to the LDA total-energy differences $E(N)-E(N-1)$ and $E(N+1)-E(N)$ (the so-called $\Delta$SCF method~\cite{dscf}), which have been shown to agree well with the experimental atomic/molecular ionization energies and electron affinities~\cite{molecule}. However, the above procedure fails for extended systems such as solids, because adding/removing electrons on an extended eigen state only gives an infinitesimal charge-density change locally, which leads to $\tilde{E}_l(s_l)=0$ and $\lambda_l=0$ (no correction). In Ref.~[\onlinecite{ceder}], the authors proposed adding/removing a finite number of electrons in the primitive cell, which improves the LDA band gaps. However, an empirical parameter is needed to adjust the amount of electrons to be changed. So far, there is no systematic and parameter-free method to use the Koopmans' theorem for correcting band gaps from molecules to solids, and the applicability of this approach is significantly limited.

In this work, we propose a new approach that extends the above correction to solids. The basic ansatz is that the energy curve $E_l(N\pm s_l)$ for removing/adding $s_l$ fractional electrons on any single-particle orbital $\phi_l$ (not necessarily eigen states) in the valence-band/conduction-band subspace should be a straight line.
We call it an extended Koopmans' theorem, as it releases the condition that $\phi_l$ must be an eigen state.
This ansatz is plausible as in the Hartree-Fock total-energy expression, when the orbital relaxations are ignored, $E_l(N\pm s_l)$ is linear with $s_l$ regardless whether $\phi_l$ is an eigen state or not, because the nonlinear self-interaction term is cancelled out between the Hartree energy and the exchange integral. In this sense, the straight-line condition removes the self-interaction energy for $\phi_l$, which may improve the band gap~\cite{SIE}.
Of course, the correctness of the ansatz can only be judged by its final results.
To introduce nonzero corrections, we need localized $\phi_l$. The Wannier functions (WF)~\cite{wf1,wf2} are most localized orbitals within the valence-band and conduction-band subspaces. Hence we will use WF as $\phi_l$ in Eq.(\ref{newE}) and (\ref{newH}). We will show that the resulting eigen energies are in excellent agreement with experiments.

The DFT calculations are performed by the PEtot code~\cite{PEtot} within the LDA. Norm-conserving pseudopotentials are used. The plane-wave energy cutoff and {\it k}-point mesh guarantee the convergence of the eigen energies within 0.01 eV. For the LDA band gap calculation, we use the primitive cell of the compound with the lattice parameters and atomic coordinates fully optimized. The spin-orbit coupling (SOC) is considered.

The WFs are calculated using the wannier90 code~\cite{wannier90}. We construct the WFs for valence bands and conduction bands separately, by projecting the eigen states onto the atomic orbitals~\cite{dv1,dv2}. The atomic orbitals are chosen based on their predominance in the valence and conduction bands. For example, for GaAs, the valence-band WFs are constructed by projections onto one As {\it s} and three As {\it p} orbitals; the conduction-band WFs are constructed by projections onto one Ga {\it s} and three Ga {\it p} orbitals. The resulting WFs are shown in the Supplemental Material Fig.~S1, which are all highly localized. We have tested the projected WFs and maximally-localized WFs, and the resulting band-gap corrections differ by less than 0.01 eV.

The next task is to calculate $\lambda_l$. We find that the screening effect from other electrons plays an important role. If we remove the WF orbital from the charge density and perform non-self-consistent calculations, the band gap corrections can be overestimated by several eV. To include the screening effect, we must calculate $E_l(N\pm s_l)$ self-consistently. We perform this calculation with spin-polarization using a supercell equal to $4\times4\times4$ times primary cell, to exclude the interactions between the WF and its images. To add/remove fractional WF $\phi_l$ in the spin-up channel, we optimize all the spin-up states orthogonal to $\phi_l$ and optimize the spin-down states in the conventional way. The energy $E_l(N\pm s_l)$ is
\begin{eqnarray}
E_l(N\pm s_l)&=& \sum_{j}\langle\varphi_j |-\frac{1}{2} \nabla^2 +V_{NL}|\varphi_j \rangle +  \alpha_l \langle\phi_l |-\frac{1}{2} \nabla^2 +V_{NL}|\phi_l \rangle + \int v_{ion}(\textbf{r}) \rho(\textbf{r}) d\textbf{r} +E_{Hxc}[\rho(\textbf{r})] \nonumber \\
\rho(\textbf{r})&=&\sum_{j} |\varphi_j|^2+\alpha_l |\phi_l|^2
\end{eqnarray}
where $V_{NL}$ is the nonlocal-potential operator, $v_{ion}(r)$ is the ionic potential, and $E_{Hxc}$ is the conventional Hartree and LDA XC energy. For valence-band WFs, $\alpha_l=1-s_l$ and the summation of $j$ is from 1 to $N/2-1$ ($N$ is an even number) for the spin-up channel, and 1 to $N/2$ for the spin-down channel; for conduction-band WFs, $\alpha_l=s_l$, and the summation is from 1 to $N/2$ for both spin-up and spin-down channels. The $\varphi_j$ in the spin-up channel is required to be orthogonal to $\phi_l$, i.e., $\langle\varphi_j|\phi_l\rangle=0$. Using a Langrangian multiplier for this constraint, the minimization of $E_l(N\pm s_l)$ with respect to $\varphi_j$ (while $\phi_l$ is kept fixed) yields
\begin{equation}
 H_{\rm LDA} |\varphi_j\rangle - \beta_j |\phi_l\rangle=\varepsilon_j |\varphi_j\rangle
\end{equation}
with $\beta_j=\langle\phi_l|H_{\rm LDA}|\varphi_j\rangle$ for the spin-up channel and $\beta_j=0$ for the spin-down channel. The conventional conjugate-gradient formalism can be used to solve the above equation and obtain the minimum $E_l(N\pm s_l)$. Note, when $s_l=0$, Eq.(3) returns to the conventional DFT expression. A few $s_l$ values need to be calculated to obtain $\lambda_l$. Using Eq.(\ref{newH}), the modified eigen energy of the LDA eigen state $\psi_i$ can be expressed as:
\begin{equation}
\epsilon_i = \epsilon_i^{\rm LDA} + \sum_l \lambda_l |\langle\phi_l|\psi_i\rangle|^2
\end{equation}
with $\epsilon_i^{\rm LDA}$ being the original LDA eigen energy. Note, one could solve Eq.(\ref{newH}) self-consistently for $\psi_i$. However, the self-consistent effect for bulk eigen states is rather small. Here we just use the original LDA wave function $\psi_i$, and take the expectation value of Eq.(\ref{newH}).

\begin{figure}[!htpb]
\includegraphics[bb=19 26 500 400, width=10cm]{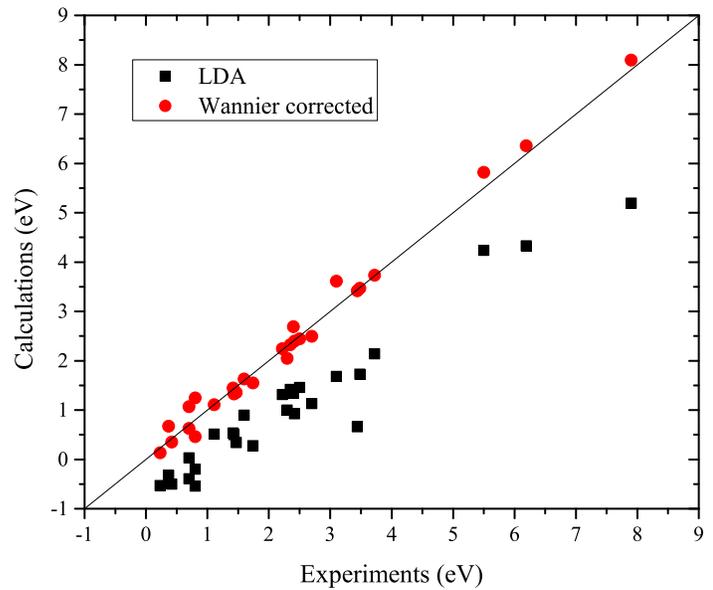}
\caption{The calculated LDA and Wannier-corrected band gaps versus the experimental band gaps~\cite{data-book} for the 27 solids.}
\label{bandgap}
\end{figure}

Using the method above, we have calculated 27 semiconductor compounds (Supplemental Material), including conventional semiconductors and oxides with experimental band gaps ranging from 0.2 to 8 eV, covering a wide range of physical situations and application interests. The Wannier-corrected band gaps along with the LDA band gaps are plotted in Fig.~\ref{bandgap} versus the experimental band gaps~\cite{data-book}. Our LDA band gaps (Table~S1) agree well with previous published results. They significantly underestimate the experimental values (even negative for some compounds), and there is no simple correlation between the LDA and experimental band gaps. The Wannier-corrected band gaps are in good agreement with experiments, and the errors are on par with more expensive methods such as the GW~\cite{GWbook}. In the following, we discuss in more details for a few examples.

GaAs is a representative of main-group semiconductors, it has a direct band gap of 1.43 eV in experiments. The LDA band gap is 0.5 eV after considering the SOC. As stated before, we construct one {\it s} and three {\it p} projected WFs at the As site for the valence bands. The VBM state is purely contributed from the As-{\it p} projected WFs, i.e. $|\langle \phi_{\rm As,\it s} | \psi_{\rm VBM} \rangle |^2=0$ and $\sum_p|\langle \phi_{\rm As,\it p} | \psi_{\rm VBM} \rangle |^2=1$, so the VBM energy correction is $\lambda_{\rm As,\it p}=-0.58$ eV. The CBM energy correction is purely from the Ga-{\it s} projected WF, i.e., $|\langle \phi_{\rm Ga,\it s} | \psi_{\rm CBM} \rangle |^2=1$, with $\lambda_{\rm Ga,\it s}=0.25$ eV. Therefore, the Wannier-corrected band gap is $0.5+0.58+0.25=1.33$ eV, which agrees well with experiments. At the bottom of the valence bands, the $\Gamma_{6v}$ ($\Gamma_{1v}$ without SOC) state is purely contributed from the As-{\it s} projected WF, and the energy correction is $\lambda_{\rm As,\it s}=-0.88$ eV, which is 0.3 eV more than the correction to the VBM. The LDA bottom valence band is 12.84 below the VBM, so the Wannier-corrected value is 13.14 eV, which agrees excellently with the experimental value of 13.1 eV~\cite{gaas1}.

ZnO is a prototypical metal oxide with a direct band gap of 3.4 eV in experiments. However, its LDA band gap is only 0.66 eV. There are controversies for whether the GW can reproduce the experimental results~\cite{zno-louie,zno-gw}. The typical hybrid functional also gives a too small band gap (2.5 eV)~\cite{zno-hse}. For the valence bands, we constructed five Zn-{\it d} projected, one O-{\it s} projected, and three O-{\it p} projected WFs. Due to the strong {\it p-d} hybridizations in ZnO, both the O-{\it p} and Zn-{\it d} projected WFs contribute to the VBM, with the energy corrections of -1.1 eV and -0.9 eV, respectively. For the CBM, similar to the GaAs case, only the Zn-{\it s} projected WF contributes, and the energy correction is $0.75$ eV. As a result, the Wannier-corrected band gap is 3.41 eV, in excellent agreement with experiments. We note that a major part of the large correction comes from the {\it d} component in the VBM. For II-VI systems with weaker {\it p-d} hybridizations, e.g. ZnS, the {\it d} projected WFs only contribute $\sim$0.2 eV to the VBM correction and the LDA band gap errors tend to be smaller than that of ZnO.

\begin{figure}[!htpb]
\includegraphics[bb=19 26 500 400, width=10cm]{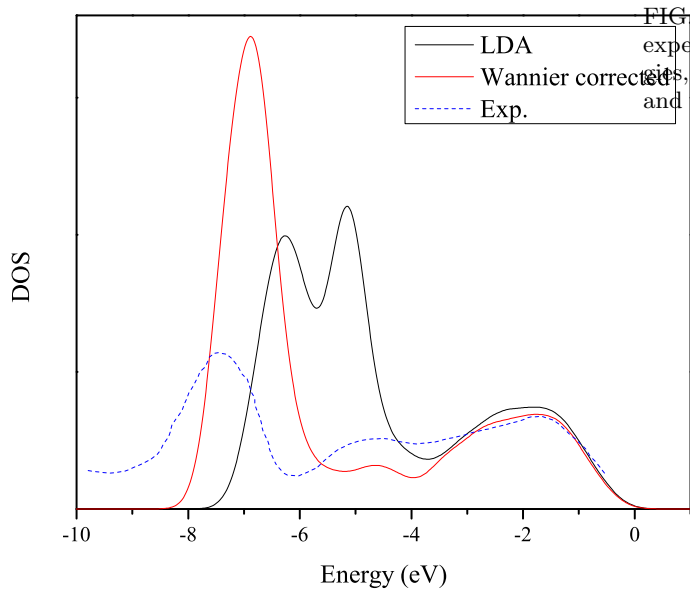}
\caption{The DOS of ZnO from the LDA and Wannier-corrected calculations. The dashed line is the experimental UPS curve taken from Ref.~[\onlinecite{zno-dos}]. Because the cross section is not considered in DOS calculations, the height of the experimental curve should not be compared with the calculations.}
\label{dos}
\end{figure}

Another interesting quantity in ZnO is the energy position of the Zn 3{\it d} bands. The density of states (DOS) calculated by the LDA and the new method along with the experimental results are plotted in Fig.~\ref{dos}. The experimental Zn 3{\it d} peak is $\sim$7.5 eV below the VBM~\cite{zno-1}. The LDA peaks are $\sim$2 eV higher compared to experiments. The Wannier-corrected peak is $\sim$7 eV below the VBM, which agrees with the GW results and is better than the hybrid functional results~\cite{zno-hse,zno-2}. Furthermore, in experiments, Zn 3{\it d} is a single peak, while LDA yields two peaks and the low-energy peak couples strongly with the O 2{\it p} states~\cite{zno-hse}. Our Wannier-corrected method reduces the {\it p-d} repulsion, and merges the two peaks into a single one in agreement with experiments.

The above discussion indicates that the Wanner-corrected method not only improves band gaps, but also improves band energies inside valence bands. The same is true for conduction bands. Table~\ref{conduc} shows that the conduction band energies at the $\Gamma$, X, and L points are all corrected well by our method, for two most widely-studied semiconductors: Si and GaAs.

The above results show that the new method works well for solids. One remaining question is whether it also works for molecules. In molecule calculations, we use the open-boundary condition for the Poisson equation to avoid image interactions~\cite{PEtot}.
For atoms or simple molecules such as LiCl, the projected WFs are just the eigen orbitals, so our method behaves the same as the $\Delta$SCF method and agrees well with experiments~\cite{molecule}.

\begin{figure}[!htpb]
\includegraphics[bb=56 0 290 439, width=10cm]{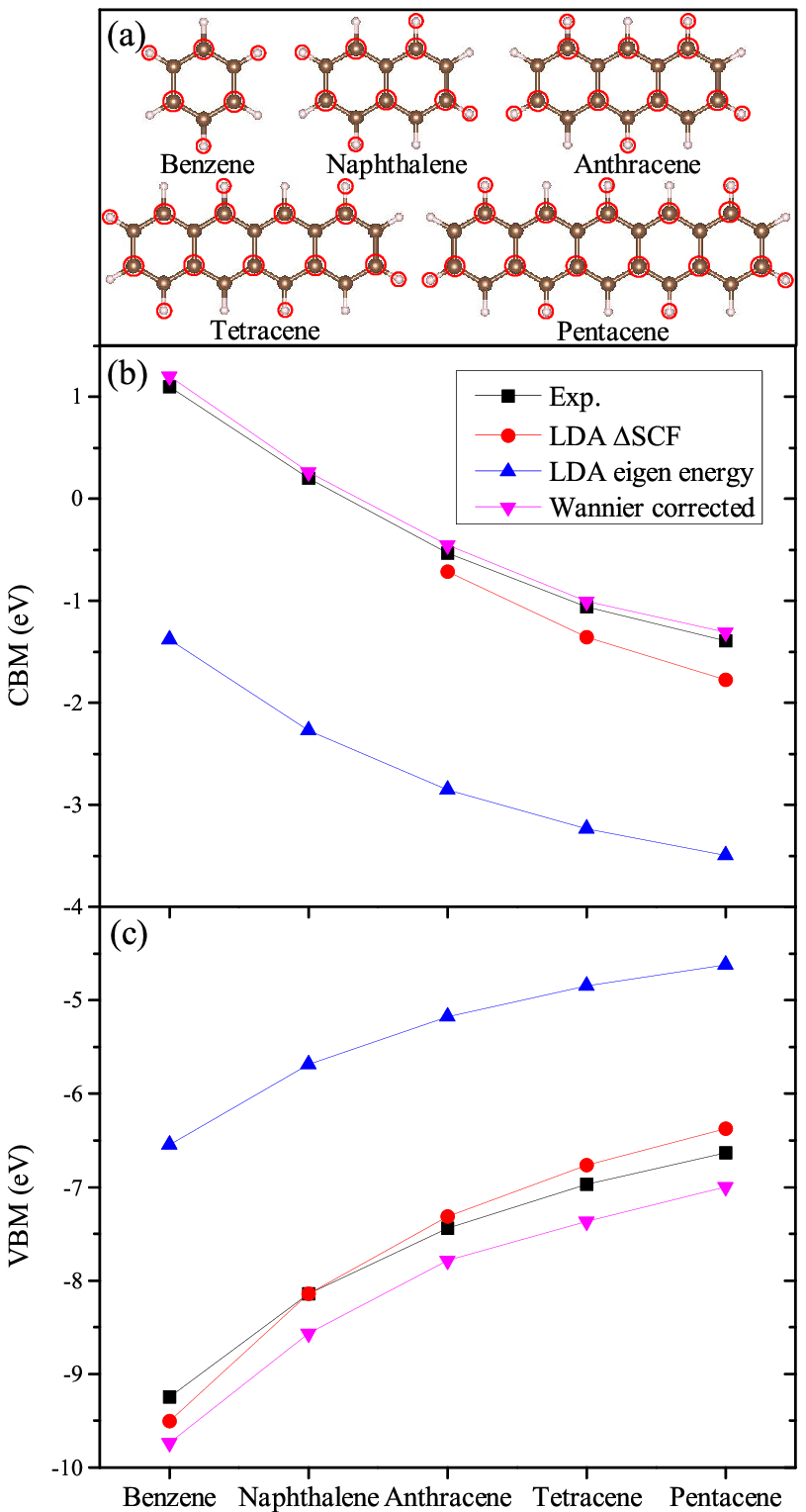}
\caption{The structures of the calculated molecules (a), the experimental energies~\cite{pah-exp}, $\Delta$SCF energies, LDA eigen energies, and Wannier-corrected eigen energies for the CBM (b) and the VBM (c).}
\label{molecule}
\end{figure}

Next we discuss a more complex molecule serial: polycyclic aromatic hydrocarbons, shown in Fig.~\ref{molecule}(a). As the number of benzene-ring increases, the molecules will eventually become a 1D system. We symmetrically pick half of the carbon and hydrogen atoms [circled by red in Fig.~\ref{molecule}(a)], and construct C-{\it s}, C-{\it p}, and H-{\it s} projected WFs on these atom sites. Other choices have been tested and the results are similar [to keep the symmetry, the WFs at the atoms uncircled in Fig.~\ref{molecule} (a) can also be included in Eq.(\ref{newH}), with an overall factor of 1/2 applied to all $\lambda_l$]. In these cases, the WFs are no longer the eigen orbitals. The VBM of these molecules are purely contributed from the the C-$p_z$ projected WFs constructed by the occupied orbitals, and the CBM are purely contributed from the C-$p_z$ projected WFs constructed by the unoccupied orbitals. Figure~\ref{molecule} plots the experimental VBM and CBM~\cite{pah-exp}, the LDA eigen energies, the $\Delta$SCF energy [$E(N)-E(N-1)$ and $E(N+1)-E(N)$], and our Wannier-corrected results. The LDA eigen energies are higher for the VBM and lower for the CBM by 2$\sim$3 eV compared to the experiments. The Wannier-corrected results agree excellently with the experiments for the CBM, and are only $\sim$0.4 eV lower than the experimental values for the VBM. Although the $\Delta$SCF energies also agree well with the experiments (note for benzene and naphthalene, the CBMs are above vacuum, so $E(N+1)$ cannot be calculated), the energy trend of the new method is much better.
As the number of benzene-ring increases, the $\Delta$SCF energy errors (compared to experiments) for both the CBM and VBM keep increasing, and eventually the $\Delta$SCF energies will converge to the LDA eigen energies for the extended 1D system (the increase of the error is nevertheless slow, which explains why the $\Delta$SCF works for moderate-sized molecules). However, the error of our method does not increases with the molecule size. It indicates that the new method works from small molecules all the way to extended solids, and thus it is a general approach applicable to all systems for electronic structure calculations.

In summary, we have presented a general and efficient method to calculate electronic structures from molecules to solids. This method is based on an extended Koopmans' theorem, and uses the WFs as localized orbitals for the fractional electron addition/removal. It does not have any adjustable parameters and is computationally much cheaper than the hybrid functionals or GW calculations. For solids, it yields accurate eigen energies not only for band-edge states but also for other states insides the bands; for molecules, it yields better trends than the traditional $\Delta$SCF method.

\begin{acknowledgments}
This material is based upon work performed by the Joint Center for Artificial Photosynthesis, a DOE Energy Innovation Hub, supported through the Office of Science of the U.S. Department of Energy under Award No. DE-SC0004993. Computations are performed using resources of the National Energy Research Scientific Computing Center (NERSC) at the LBNL. Wang is also supported through the Theory of Materials Project at the LBNL by the Basic Energy Science, Material Science and Engineering, Office of Science of the U.S. Department of Energy under Contracts No. DE-AC02-05CH11231.
\end{acknowledgments}


\begin{thebibliography}{42}
\bibitem[{koo()}]{koopman-note}
\bibinfo{note}{The original Koopmans' theorem only states that in the Hartree-Fock,
  the first ionization energy is equal to the HOMO energy if the orbital
  relaxations are ignored. Here following the terminology in literatures, by
  the Koopmans' theorem, we mean the straight-line behavior of the total energy
  as a function of the fractional number of electrons.}
\end{thebibliography}

\pagebreak

\begin{table}[!htpb]
\caption{The band gaps (eV) measured from the VBM to the conduction bands at the $\Gamma$, X, and L points of Si and GaAs.}\label{conduc}
\begin{ruledtabular}
\begin{tabular}{c  c |c c c }
Compound& Conduction band & LDA  &  Wannier-corrected  &  Experiment \\
\hline
   & $\Gamma_{\rm 15c}$ & 2.58 & 3.21 & 3.34 \\
   & $\Gamma_{\rm 2'c}$ & 3.53 & 4.09 & 4.15 \\
Si & X$_{\rm 1 c}$ & 0.63 & 1.2 & 1.13 \\
   & L$_{1 \rm c}$ & 1.58 & 2.17 & 2.04 \\
   & L$_{3 \rm c}$ & 3.33 & 3.96 & 3.91 \\
\hline
   & $\Gamma_{\rm 1c}$  & 0.5 & 1.33 & 1.43 \\
   & $\Gamma_{\rm15c}$ & 3.82 & 4.67 & 4.72 \\
GaAs\footnote{In Ref.[\onlinecite{gaas1}], the single group notation was used, so we average the SO-splitted energies of the $\Gamma_{\rm 15c}$ state.} & X$_{\rm1c}$ & 1.32 & 2.14 & 2.18 \\
   & X$_{\rm3c}$ & 1.52 & 2.33 & 2.58 \\
   & L$_{\rm1c}$ & 0.90 & 1.73 & 1.85 \\
\end{tabular}
\end{ruledtabular}
\end{table}

\pagebreak

\pagebreak

\pagebreak

\end{document}